\documentclass[12pt, conference]{IEEEtran} 
\IEEEoverridecommandlockouts
\usepackage{cite}
\usepackage{amsmath,amssymb,amsfonts}
\usepackage{algorithmic}
\usepackage{graphicx}
\usepackage{multicol}
\usepackage{textcomp}
\usepackage{xcolor}
\usepackage{csquotes}
\usepackage{url}
\usepackage{blindtext}

\AtBeginEnvironment{quote}{\par\singlespacing\small}
\def\BibTeX{{\rm B\kern-.05em{\sc i\kern-.025em b}\kern-.08em
    T\kern-.1667em\lower.7ex\hbox{E}\kern-.125emX}}

\begin{document}

\title{Approximate Bayesian Computation As An Informed Fuzzing-Inference System}

\author{
    \IEEEauthorblockN{Chris Vaisnor}
    \IEEEauthorblockN{Graduate Student - Artificial Intelligence M.S.} 
    \IEEEauthorblockN{cvaisno1@jh.edu}\\
    \IEEEauthorblockN{\textit{Johns Hopkins University}}
    \IEEEauthorblockN{\textit{Baltimore, MD}}
}
\maketitle

\begin{abstract}
The power of fuzz testing lies in its random, often brute-force, generation and execution of inputs to trigger unexpected behaviors and vulnerabilities in software applications. However, given the reality of infinite possible input sequences, pursuing all test combinations would not only be computationally expensive, but practically impossible. Approximate Bayesian Computation (ABC), a form of Bayesian simulation, represents a novel, probabilistic approach to addressing this problem. The parameter space for working with these types of problems is effectively infinite, and the application of these techniques is untested in relevant literature. We use a relaxed, manual implementation of two ABC methods, a Sequential Monte Carlo (SMC) simulation, and a Markov Chain Monte Carlo (MCMC) simulation. We found promising results with the SMC posterior and mixed results with MCMC posterior distributions on our white-box fuzz-test function.\footnote[1]{Project code can be found at: \url{https://github.com/cvaisnor/fuzzy_inference_with_ABC}}

\end{abstract}

\section{Introduction}
In the domain of cyber-security, fuzz testing or "fuzzing" is the task of testing software for security vulnerabilities based on input data. In most software applications, user input is passed in and used as part of the computation process. Malicious users, commonly referred to as "bad actors", will try and pass incorrect, broken, or unexpected values into the program to cause a bug to learn how the program operates. Perhaps the bad actor has access to the hardware the program is running on and intercepts logs from Random Access Memory (RAM), or the program is running on a business's cloud server and the actor wants to learn if program calls are being made over the internet to other service providers. 

There are several methods as well as specific software that help to automate the process for the developer writing the program that eventually gets tested. With the source code and full control of the program, an infinite number of tests can be ran with many different variations. Without the source code or limited or restricted access to the software, attempts to break the system should be considered accurate, comprehensive, and with as minimal attempts as possible. 

One such method that has not been explored is using Approximate Bayesian Computation (ABC) to infer posterior distributions of fuzz parameters. ABC falls under a class of approximate computational methods based on Bayesian statistics. \cite{sunnaker2013approximate} The fundamental task of ABC is to be able to identify parameter configurations that allow a model to generate synthetic data, that is sufficiently similar to actual data. Generally, the prior beliefs of a simulation model are updated using a likelihood function, but for certain networks, the likelihood function is not explicitly computable. ABC is a class of algorithms that are sometimes referred to as "Likelihood-free Inference".

In contrast to running a comprehensive test of all possible inputs, ABC could be used to dynamically adjust our fuzz testing strategy, focusing on high-risk areas, resulting in more efficient identification of potential vulnerabilities or breakdown sequences, and improving the cost-effectiveness of the fuzzing process. This paper provides an experiment and analysis of two of these computation methods, Sequential Monte Carlo (SMC) and Markov Chain Monte Carlo (MCMC), for creating an input sequence for fuzz testing a program as an alternative to brute force methods.

\section{Algorithm Synopsis}
Sequential Monte Carlo (SMC) and Markov Chain Monte Carlo (MCMC) simulations are probabilistic methods used to generate samples from a complex distribution. In their traditional form, they are a cornerstone of statistical methodologies, especially in Bayesian inferences where we need to work with the posterior distribution which can be an intrinsically complex high-dimensional object. 

SMC approaches sampling by considering a sequence of incrementally evolving distributions with transitional dynamics, starting with a simple prior and ending with the desired complex posterior. It generates a set of hypothetical states, or particles, from the prior distribution, and then iteratively transitions these particles to capture the evolving distributions. The transitions are accompanied by a re-weighting of the particles according to a likelihood function that reflects the fitness of a particle.

MCMC, on the other hand, navigates sampling by constructing a Markov Chain whose stationary distribution is the desired complex distribution. It starts with an initial state and uses a transition mechanism (proposal function) to propose new states. Each proposed state is either accepted (with a probability computed using an acceptance rule) or rejected. Over many iterations, the Markov Chain converges to the desired distribution.

In our research, we built manual implementations of SMC and MCMC. Although these simulations aim to circumvent the need for an explicit likelihood function, in this case, we purposely implemented a likelihood function as a proof of concept of how particles can be directed for a fuzz test. The likelihood function governs both the re-weighting of particles in the SMC and the acceptance probability of proposed states in the MCMC. We have additionally introduced a unique twist to the likelihood by considering the distance of particles/states from a target point and a penalty term for deviating from zero in the first dimension, which offers a controlled bias in the search of states. This directed controlling of states helps with getting non-zero outputs from our fuzz test function and keeping the simulation within reasonable computing standards.

\begin{figure*} [!t]
\centering
\includegraphics[width=0.9\textwidth]{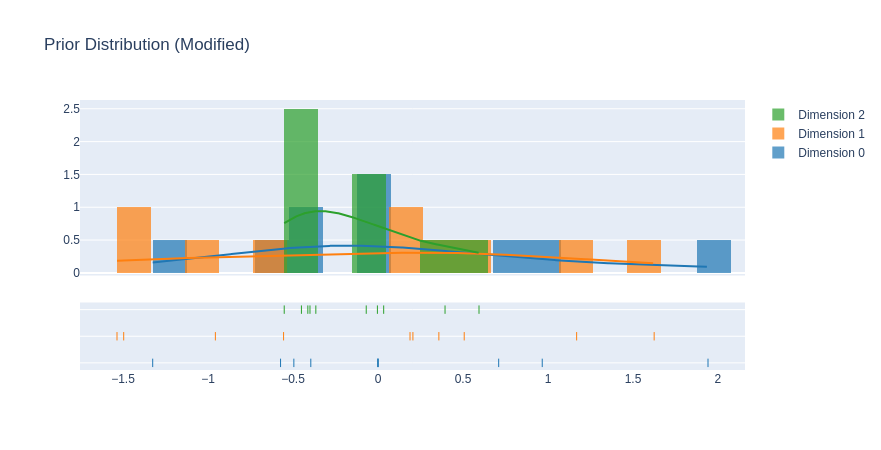}
\caption{Distribution of prior particles with the modified slice.}
\label{fig:prior_distribution_modified.png}
\end{figure*}

\section{Literature Review}
There are two main difficulties of ABC as noted in \cite{10.1093/sysbio/syw077}. The first issue is efficiently finding the correct parameter values for a given model. The second issue is defining the similarity between synthetic (simulator) data and real data.

In layman's terms, how do we get the model to output data that is accurate enough that it could be real? And how do you know what real data even looks like? You use a probability distribution over the outputs of the model. This touches on the core of Bayesian statistics, where prior probabilities are updated given new evidence. 

The domain of fuzz testing, or "fuzzing," has been subject to extensive research and development. Over the years, analyst and developer communities have developed a diverse set of techniques and tools to facilitate this process.

There are three general categories that fuzz tests can fall into, black-, grey,- and white-box fuzzing. \cite{9166552} Black-box refers to fuzz tests that are performed when the user has no prior knowledge of the software.  There are two classes of techniques currently used. One technique is a mutational iteration method where random values in the input are changed until the test period completes. The other technique is a brute-force generational method where inputs are made from scratch.  The apparent benefit is the lack of prerequisite insight into the software. Yet, its associated drawback is an often high number of impractical tests which lower the efficiency of the overall testing routine. \cite{9166552}

Gray-box is a middle ground where the user is partially aware of program logic and can use feedback to steer the direction and types of fuzz tests.  This strikes a balance between ease of use and effectiveness, therefore providing a higher chance of uncovering vulnerabilities and reaching deeper into the code. Tools like American Fuzzy Lop (AFL), LibFuzzer, and Honggfuzz are notable examples that leverage such methods. \cite{libFuzzer}

The last general category is white-box fuzzing where the user has a full understanding of the source code and can take advantage of symbolic execution to find interesting program paths. These programs tend to use "Constraint-Solvers" to accurately direct input. One such method is to incorporate Satisfiability Modulo Theory (SMT). This method relies on formal logic to test if a function is satisfiable. \cite{10.1007/11513988_4} Microsoft has credited the white-box tool SAGE with saving potentially millions of dollars on the development of Windows 7. \cite{9166552}

In recent years, more teams have been incorporating aspects of machine learning and probabilistic modeling to enhance the effectiveness and efficiency of fuzz testing. These include "Grammar-based Fuzzing," where the fuzz testing is directed by a formal grammar, and "Evolutionary Fuzzing," where genetic algorithms are used to evolve test cases. Tools that exemplify these advanced techniques include libFuzzer and honggfuzz. \cite{9166552}

It is the responsibility of the analyst or developer to assess the most effective approach depending on their needs. The growing interest in advanced probabilistic models highlights the increasing complexity of software systems and the resultant need for even more sophisticated testing techniques.

\section{Methodology}
Our research aims to generate synthetic posterior particles that provoke the same response from a "fuzz test function" as the original prior particles. We start by elucidating the user configurations, and clearly defining the constraints applicable to our experiment, such as parameters of the prior distribution and time steps for the simulation.

To assemble the prior sequences, a PyTorch tensor \cite{NEURIPS2019_9015} will be used consisting of continuous floating point values. Each particle is one row in the tensor, and each dimension of that row is a sample from a Gaussian distribution. A small slice (30 percent) of this prior is then modified such that the first dimension is hard-coded to a value of zero. This slice represents the particles that will pass our white-box fuzz test function. See Fig. \ref{fig:prior_distribution_modified.png}.

Both the number of initial particles, as well as their dimensionality, are configurable. We used 10 prior particles, with 100 dimensions each. One-third of these particles then have their first dimension set to zero. As an initial test, the prior particles are passed to the fuzz test function where a percentage indicates the proportion of particles that passed the test. The fuzz test function is a simple counter where the first dimension of each particle is checked to be within a range of -0.5 and 0.5. The higher the prior standard deviation, the closer the fuzz test returns 30 percent of particles as passing due to the possible distance that dimension is from zero.

\begin{figure} [ht]
    \centering
    \includegraphics[width=0.9\linewidth]{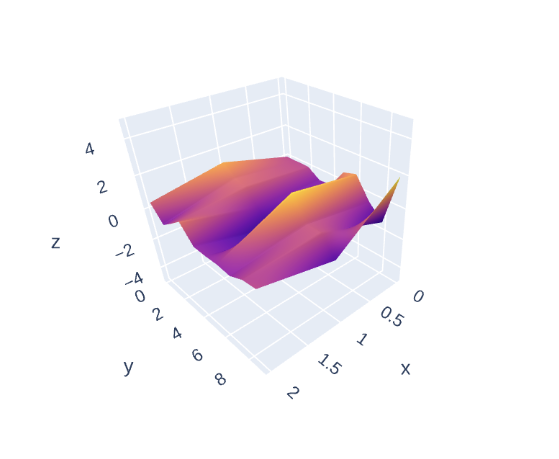}
    \caption{Surface plot of prior dimensions[1-3].}
    \label{fig:prior_surface}
\end{figure}

The prior is then passed to a manual implementation of SMC and MCMC. For both simulation methods, the functionality of transition and likelihood are created as separate helper functions. For the likelihood function, SMC uses Euclidean distance as well as an Alpha parameter to control bias in the first dimension. MCMC uses an Accept-Reject method that incorporates the same likelihood function. Each simulation is then run with 1000 time steps, generating 1000 posterior particles. 

The convergence of each simulation is then visualized with interactive plots. For SMC, the sum of weights is used as the indicator, with smaller weight updates representing converging. For MCMC, a trace plot is used. After each simulation finishes, the posterior outputs are passed to the fuzz test function to view the proportion of particles that passed the test.

\section{Results}
Our experiment falls under the white-box fuzz test category because we manually configured the test function to have certain criteria. Given our prior parameters: 100 dimensions, 10 initial prior particles, a prior mean of zero, a standard deviation of 10, and 1000 time steps, SMC was extremely successful in replicating the passing particle distributions. According to our fuzz test, 30 percent of the prior particles passed, this is due to our manual modification of the points during the generation phase. The posterior particles (count of 1000) from the SMC algorithm passed with a proportion of 89.7 percent. This shows that our likelihood function and weight-updating methods were able to steer the new particles in the "passing" direction. Although we used an explicit likelihood function, which is not standard under normal SMC conditions, this does provide a minimally viable test case.

Our MCMC algorithm did not share the same success as SMC. Due to the initialization of the algorithm, only one particle can be used to start the generation process in our single-chain configuration. We used various time steps (1000-5000) and various burn-in counts (100-500) and found mixed results. Viewing the trace plot for this algorithm also shows that the suggested value at dimension zero varied substantially, indicating a lack of convergence. The fuzz test results showed a passing rate of 23 percent on the synthetic posterior particles.

\section{Discussion and Reflections}
Due to the subject of this paper being a proof of concept, several methodological changes could be implemented to test the simulations more strenuously. The use of a white-box style testing function biases and informs the user of the correct fuzzing input. Ideally, a separate party could set up a grey or black-box function, and the simulation user would have to test different combinations and types of inputs to learn the logic of the function. 

\begin{figure*} [!t]
    \centering
    \includegraphics[width=0.9\linewidth]{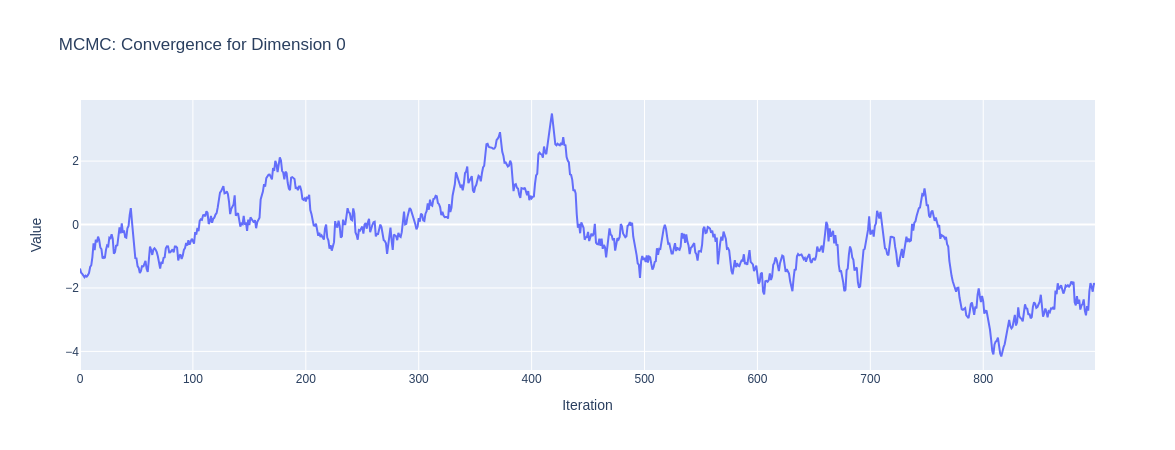}
    \caption{Markov Chain Monte Carlo (MCMC) Trace Plot}
    \label{fig:mcmc_trace}
\end{figure*}

In this experiment, the exact logic of the fuzz test function plays a critical role in determining success with the specific set of priors. It also relies on having at least one prior particle that already clarifies one output of the fuzz function. This is analogous to trying to determine the bias of a coin without ever having flipped it. Changing the testing order could be one form of alternate setup. For each time step, a batch of posterior particles could be tested against the fuzz function, and then the likelihood could be inferred. Or perhaps directly, the likelihood function becomes the fuzz test function. This still runs into the problem of setting an initial passing-particle, otherwise, the task transforms into another brute-force method.

It is these initial parameters, coupled with the likelihood evaluation, that seem to decide the success of generating a posterior that passes the test. Under more strict testing circumstances and without an idea of what priors could work, it's possible that these generation methods are not the best for this type of input. There are other topics even within the fuzz testing category where these ABC methods could be used. One such category could be generating alternate execution paths given a set of parameters as noted in \cite{10.1007/978-3-540-78800-3_27}. Rather than a single-function fuzz test where samples are generated from a prior with fixed size-and-dimension floating-point values, the prior would be the execution path or call stack for an entire program. This execution path then begins to look like a traditional Bayesian network with each random variable denoting one direction the program could take. 

\section{Conclusion and Future Work}
As an initial investigation, our results highlight the potential of applying Approximate Bayesian Computation (ABC) to the problem of fuzz testing. As made clear in the discussion section, there are many different ways to test these ABC methods and this paper only covers a small section of the fuzz testing problem set. Future work could investigate other aspects of cyber-security as well, such as particle-based methods for hash breaking or probing wireless network topography. Given appropriate time and resources, a deeper analysis could be conducted on the computational complexity of current methods versus Bayesian probability simulations. Overall, our avant-garde approach to fuzz testing hopes to set the stage for further research.

\bibliographystyle{ieeetr}
\bibliography{References}

\end{document}